\documentclass[twoside,english,preprint]{elsarticle}
\usepackage[T1]{fontenc}
\usepackage[latin9]{inputenc}
\usepackage{graphicx}

\makeatletter
\journal{Solid State Communications}

\usepackage{babel}

\makeatother

\usepackage{babel}
\begin{document}

\title{Unexpected large thermal rectification in asymmetric grain boundary
of graphene}

\author[rvt]{Hai-Yuan Cao}

\author[rvt]{Hongjun Xiang}

\author[rvt]{Xin-Gao Gong}

\ead{xggong@fudan.edu.cn}

\address[rvt]{Key Laboratory of Computational Physical Sciences (Ministry of Education),
State Key Laboratory of Surface Physics, and Department of Physics,
Fudan University, Shanghai 200433, P.R. China}
\begin{abstract}
We have investigated the lattice thermal transport across the asymmetric
tilt grain boundary between armchair and zigzag graphene by nonequilibrium
molecular dynamics (NEMD). We have observed significant temperature
drop and ultra-low temperature-dependent thermal boundary resistance.
More importantly, we find an unexpected thermal rectification phenomenon.
The thermal conductivity and Kapitza conductance is direction-dependent.
The effect of thermal rectification could be amplified by increasing
the difference of temperature imposed on two sides. Our results propose
a promising kind of thermal rectifier and phonon diodes based on polycrystalline
graphene without delicate manipulation of the atomic structure. \end{abstract}
\maketitle
Graphene, a two dimensional honeycomb lattice, is well recognized
as one of the most promising materials in future electronics and nanotechnology\citep{1,2}.
Numerous previous studies have shown that the monolayer graphene samples
with perfect lattice structure exhibit outstanding electronic, thermal,
mechanical and optical properties\citep{3,4,5,6}. While the size
of graphene flakes from mechanical exfoliated is too small to be utilized
for mass production of functional devices, chemical vapor deposition
(CVD) enables the synthesis of large-area, high-quality and low-cost
graphene sheets\citep{7,8,9}. Due to the imperfections of metal foils
in the growth process, the grains inevitably nucleate during the growth
which results to the formation of grain boundaries in graphene sheets\citep{10,11}.
Understanding the effect of the grain boundaries on the mechanical,
electronic, and thermal properties is very helpful and necessary for
future applications of the larger-scale polycrystalline graphene.

Recently, several studies have shown that the grain boundaries have
crucial influence on the mechanical and electronic properties. For
the mechanical property, the grain boundaries with high density of
topological defects show anomalous characteristics of mechanical strength\citep{12}.
In the case of charge transportation, the charge mobility depends
on the grain size which is determined by the grain boundaries\citep{9}.
Furthermore, the transmission of the charge carrier across the grain
boundary depends on the symmetry of the atomic structure. The charge
transport of the symmetric grain boundary reveals high transparency
while that of the asymmetric grain boundary emerges almost perfect
reflection\citep{13}. Experiments have shown that the line defects
like grain boundary can be one of the most powerful ways to manipulate
electronic properties at nanoscale\citep{14,15,16}. For the thermal
transport, although there have been many studies on the point defects
including vacancies\citep{17}, Stone-Wales defects\citep{18} and
isotope defects\citep{19}, the effect of line defects like grain
boundary on thermal conductivity is still far from well-understood.
In a recent work \citep{20}, thermal transport across the tilt grain
boundaries with simplest symmetric atomic structure has been considered.
Sudden temperature drop at the grain boundary and superior boundary
conductance has been observed in that system\citep{20}. There is
still a long way for us to further explore and understand the impact
of one-dimensional global defects on the thermal transport of graphene.
Similar to the charge transport, it is reasonable to expect that the
asymmetric grain boundary would exhibit different property of thermal
transport from that of the symmetric grain boundary.

In this paper, we adopt the nonequilibrium molecular dynamics (NEMD)
method\citep{21,22,23} to simulate the heat transport across the
asymmetric grain boundary between armchair graphene and zigzag graphene.
We show that the Kapitza conductance and the thermal conductivity
have strong dependence on the temperature. Furthermore, surprising
thermal rectification phenomenon has been observed in the graphene
with such asymmetric grain boundary. Our results provide a practical
choice for designing thermal rectifier in polycrystalline graphene
without delicate atomic engineering.

Figure 1 shows the atomic structure of the asymmetric grain boundary
connecting the armchair and the zigzag regions. This structure has
been denoted as a boundary structure between the armchair graphene
and the zigzag graphene\citep{24} while the electronic transport
across that has been studied \citep{13,25}. This structure is composed
of a periodic array of 5-pentagon and 7-heptagon topological defects.
According to the mismatch of the period of the zigzag part and the
armchair part, this asymmetric grain boundary shows peculiar 7-5-7
structure while the standard pentagon-heptagon dislocation cores would
flip their orientations from 5-7 to 7-5\citep{24}. In spite of the
lattice mismatch, the formation energy of this asymmetric grain boundary
is much smaller than the typical energy of the bare edges in graphene\citep{26}.

In all simulations, velocity Verlet algorithm is used to integrate
the equation of motions\citep{27}. Periodic boundary condition is
imposed on the direction parallel to the grain boundary. Along the
direction perpendicular to the grain boundary, fixed boundary condition
is applied on the green atoms in the Figure 1. The brown atoms in
Figure 1 are coupled to $Nos\acute{e}-Hoover$ heat baths\citep{28},
whose temperatures are set to $T_{high}$ and $T_{low}$, respectively,
to generate the temperature gradient. \textbf{The length of heat bath used in our simulation is around 1 nm. We have carefully tested the effect of the heat bath length on our results. The quantity of the calculated thermal rectification is only slightly affected by the length of heat bath.} A recent parameter-modified
Tersoff potential is used to describe the atomic interaction\citep{29},
which has been verified to give the right velocities of the acoustic-phonon
which is crucial in thermal transport. The new parametrized potential
considerably improves the computational values of the lattice thermal
conductivity and makes them in good agreement with the experimental
results\citep{18,30}. After the system reaches to the steady state,
the heat flux could be calculated by the energy flowing from the heat
bath to the system\citep{31}. The thermal conductivity could be obtained
according to the Fourier's law 
\begin{equation}
\kappa=\frac{Jd}{\Delta T_{1}wh}
\end{equation}
 In the equation, $J$ represents the heat flux, $d$ is the length
of sample, $\Delta T_{1}$ is the temperature difference between $T_{high}$
and $T_{low}$, $w$ and $h$ is the width and height of the monolayer
graphene, respectively. The temperature usually shows significant
temperature jump at the grain boundary region, the boundary conductance
is defined as: 
\begin{equation}
G=\frac{J}{\Delta T_{2}}
\end{equation}
 $J$ is the heat flux and $\Delta T_{2}$ is the temperature jump
at the boundary\citep{20}.

Before starting the simulation, the atomic structure has been fully
optimized in a microcanonical ensemble by conjugate gradient method
for ensuring the forces on each atom being less than $10^{-8}$ eV/$\mathring{A}$.
Applying the $Nos\acute{e}-Hoover$ thermostat on two ends about $1\ ns$,
the system reaches steady equilibrium.\textbf{ }The heat flux and
the temperature gradient is obtained by the average of $4$ ns. The
time step of the simulation is set to 1 fs. The simulation time has
been tested to be enough for convergence. The width of the period
parrallel to the grain boundary is about $3$ nm and the height of
the sample is set to be the distance between graphite layers about
$3.35\ \mathring{A}$\citep{20}.

At first, we calculate the thermal conductivity of pure graphene to
validate our simulation results. While the length of the sample is
about 10 nm and 25 nm under the simulation temperature around 300
K, the thermal conductivity is about 327 W/(mK) and 455 W/(mK) respectively,
which is in good agreement with the previous NEMD simulation results
\citep{32,33,34}. Under the same simulation condition, the thermal
conductivity of the graphene sheet with the asymmetric grain boundary
reduced to only around 240 W/(mK) which becomes considerably lower
than the pure graphene. The apparent reduction of thermal conductivity
is attributed to the strong phonon scattering at grain boundary.

Superior boundary thermal conductance has been observed in the simulation.
At room temperature, the boundary thermal conductance of the asymmetric
grain boundary is about $1.5\times10^{10}$ W/m$^{2}$K which is much
larger than the reported value of grain boundaries in nanocrystalline
diamond thin films\citep{35} and silicon-sillicon (001) $\Sigma29$
grain boundaries\citep{36,37}. Our results is slightly lower than
the previous reported boundary thermal conductance of the symmetric
grain boundary\citep{20}. The misorientation angle of this asymmetric
grain boundary is 30 degree which is larger than that of the reported
symmetric grain boundaries. The large misorientation angle would induce
the increase of the defects densities and the probability of the phonon-boundary
scattering, which would finally reduce the thermal boundary conductance.

The temperature dependence of the boundary thermal conductance and
the thermal conductivity has been investigated systematically, which
are shown in Figure 2(a) and 2(b), respectively. When the temperature
increases up to 550 K, the thermal boundary conductance increases
from $1.3\times10^{10}$ W/m$^{2}$K to $2.0\times10^{10}$ W/m$^{2}$K.
The similar trend has also been observed in the Si-Ge interfaces\citep{38}
and the Kr-Ar interfaces\citep{39}. While at the high temperature
around 600K, the boundary thermal conductance suddenly drops. This
could be attributed to the transformation of atomic structure of the
grain boundary. The C-C bonds in the array of 5-pentagon and 7-heptagon
topological defects broken has been observed which reveals the structural
instability of the asymmetric grain boundary at high temperature.
The thermal transport depending on the atomic structure of the grain
boundary was also found in a previous study \citep{20}. The thermal
conductivity follows the same relation as the boundary thermal conductance
respecting to the temperature, which has been reported in previous
theoretical study on defect free graphene\citep{17,40}, while the
sudden drop of the thermal conductivity near 600 K could be also attributed
to the atomic structural changing of the grain boundary.

The monotonic increase of thermal boundary conductance with the rising
of temperature before the structure changing could be interpreted
by considering the following two factors. Firstly, the thermal fluctuation
increases with the rise in temperature. Consequently, the probability
of inelastic phonon scattering increases at the grain boundaries\citep{38}.
The optical phonon with high frequency might break down into numerous
acoustic phonons with low frequency. These acoustic phonons have higher
probability of transmission comparing to high frequency optical phonon
due to the limited width of the asymmetric grain boundary. Besides
that, more acoustic phonon would be stimulated at high temperature
which could also contribute to the increase of boundary thermal conductance.

It is interesting that we find the thermal conductivity and Kapitza
conductance is asymmetric, i.e., the heat flux from armchair to zigzag
is larger than that from zigzag to armchair, which suggests that such
asymmetric grain boundary structure could be used to design a new
kind of thermal rectifier.

The thermal rectification is found to be strongly temperature dependent
and it could reach to as high as 74\%. The thermal rectification versus
the temperature difference $|\Delta T|$ and the average temperature
$T$ are plotted in Figure 3(a) and 3(b). In Figure 3(a), the temperature
difference between $T_{high}$ and $T_{low}$ has been set to 120
K and the simulation temperature varies from 200 K to 700 K. With
the decrease of the simulation temperature, the thermal rectification
increases from 1.05 to 1.73. In Figure 3(b), the average simulation
temperature $\frac{T_{low}+T_{high}}{2}$ has been set to 400 K and
the temperature difference between $T_{high}$ and $T_{low}$ varies
from 40 K to 140 K. With the increasing of the temperature difference,
the thermal rectification increase from 1.05 to 1.33. In both Figure
3(a) and 3(b), the thermal rectification increases with the rise in
the temperature difference. The nanostructure like the carbon nanotube
intramolecular junction\citep{31}, the asymmetric graphene ribbon\citep{43},
the carbon nanocone\citep{44} and the hybrid graphene-graphane nanoribbon\citep{45}
has been proposed to be thermal rectifier in the past few years. The
effect of thermal rectification in the asymmetric grain boundary revealed
in this work is comparable to those in previous proposed systems.

\textbf{Furthermore, the thermal rectification is actually determined
by the change of the phonon spectral overlap under opposite bias temperature{[}46{]}.
In fact, the obvious difference between the phonon power spectrum
of the armchair and zigzag graphene was verified in the previous theoretical
studies {[}47, 48{]}. Besides that, the temperature-dependent phonon
power spectrum, a typical characteristic of the nonlinear system,
would change the phonon spectral overlap while the temperature bias
is reversed, which results in the emerging of thermal rectification.}
On the one hand, if the armchair part is under the high temperature,
the power spectrum between these two parts is well matched and the
thermal transport could be efficiently exchanged between two parts.
On the other hand, if the armchair part is under the low temperature,
the power spectrum of the armchair and the zigzag parts are shifted
and the overlap is expected to be diminished. By the decrease of overlapping
phonon spectrum, the efficiency of the thermal exchange between the
armchair part and the zigzag part declines dramatically. While enlarging
the temperature difference in two ends, the mismatch between the phonon
spectrum of two parts would also increase therefore the thermal rectification
is magnified. Under low temperature, the thermal conductivity of the
graphene highly depends on the temperature\citep{40}, which indicates
that the phonon spectrum shifted more considerably at low temperature
than at high temperature by the same amount change of temperature.
For this reason, with the same temperature difference, the overlap
of phonon spectrum decreases more at low temperature than that at
high temperature, therefore the effect of thermal rectification is
amplified at low temperature. According to the phonon spectrum match/mismatch
theory mentioned above, the results of our simulation could be well
understood.\textbf{ On the other hand, at the high temperature region
more acoustic phonons would generate in the system. If more acoustic
phonons are generated, the grain boundary with limited width would
impose less impact on the thermal transport. In this case, the thermal
rectification would also decrease at the high temperature region. }

In summary, we have found that the Kapitza conductance across the
grain boundary and the thermal conductivity are strongly temperature-dependent.
Surprisingly, the asymmetric heat flux across the asymmetric zigzag-armchair
grain boundary is observed. It provides a potential candidate for
future design of carbon-based thermal rectifier which has attracted
a lot of attention recently \citep{49}.

\textbf{Acknowledgement}

This work is supported by NSF of China, the Special Funds for Major
State Basic Research, the Research Program of Shanghai municipality.

\clearpage{} 

\clearpage{}

\begin{figure}
\includegraphics[bb=30bp 0bp 360bp 278bp,scale=0.7]{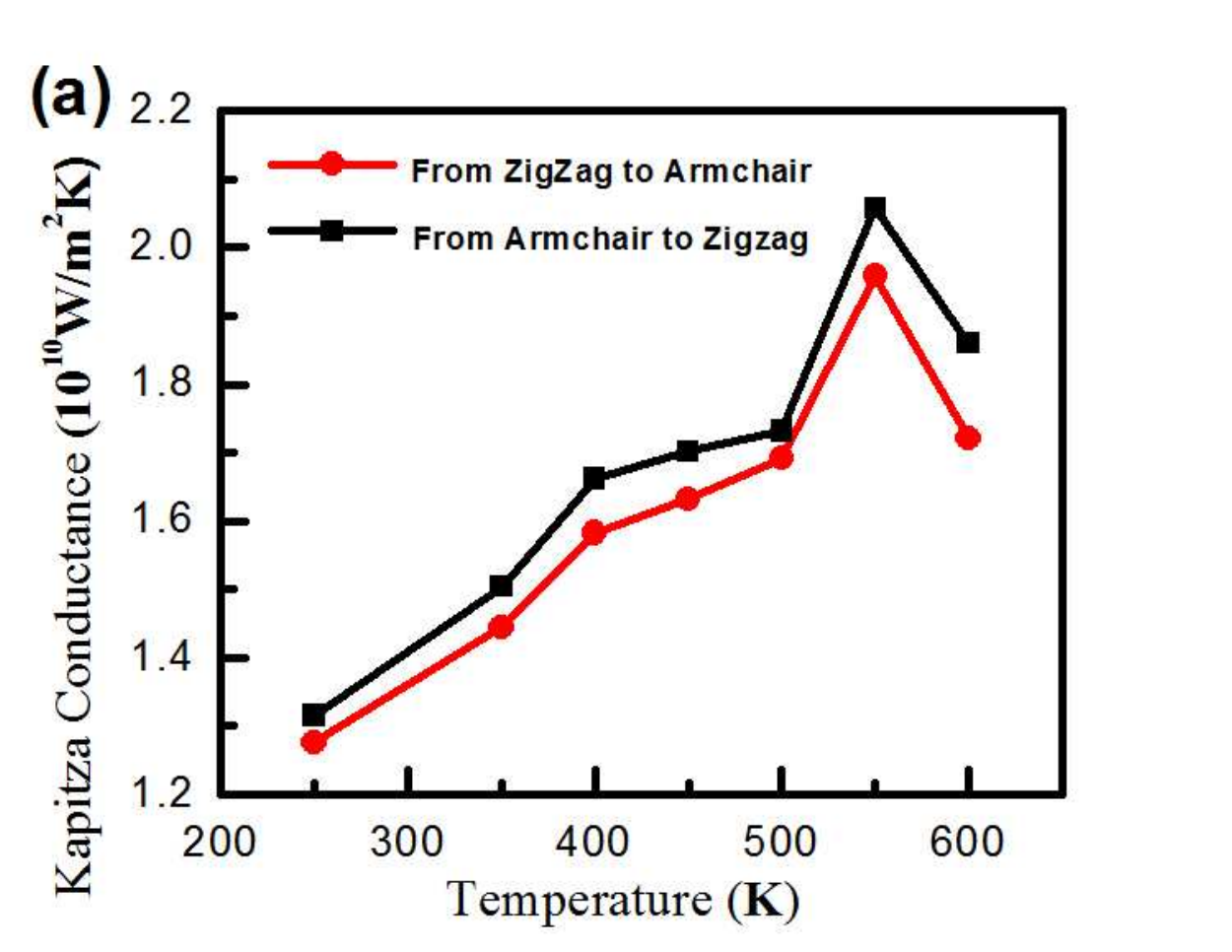}\includegraphics[scale=0.7]{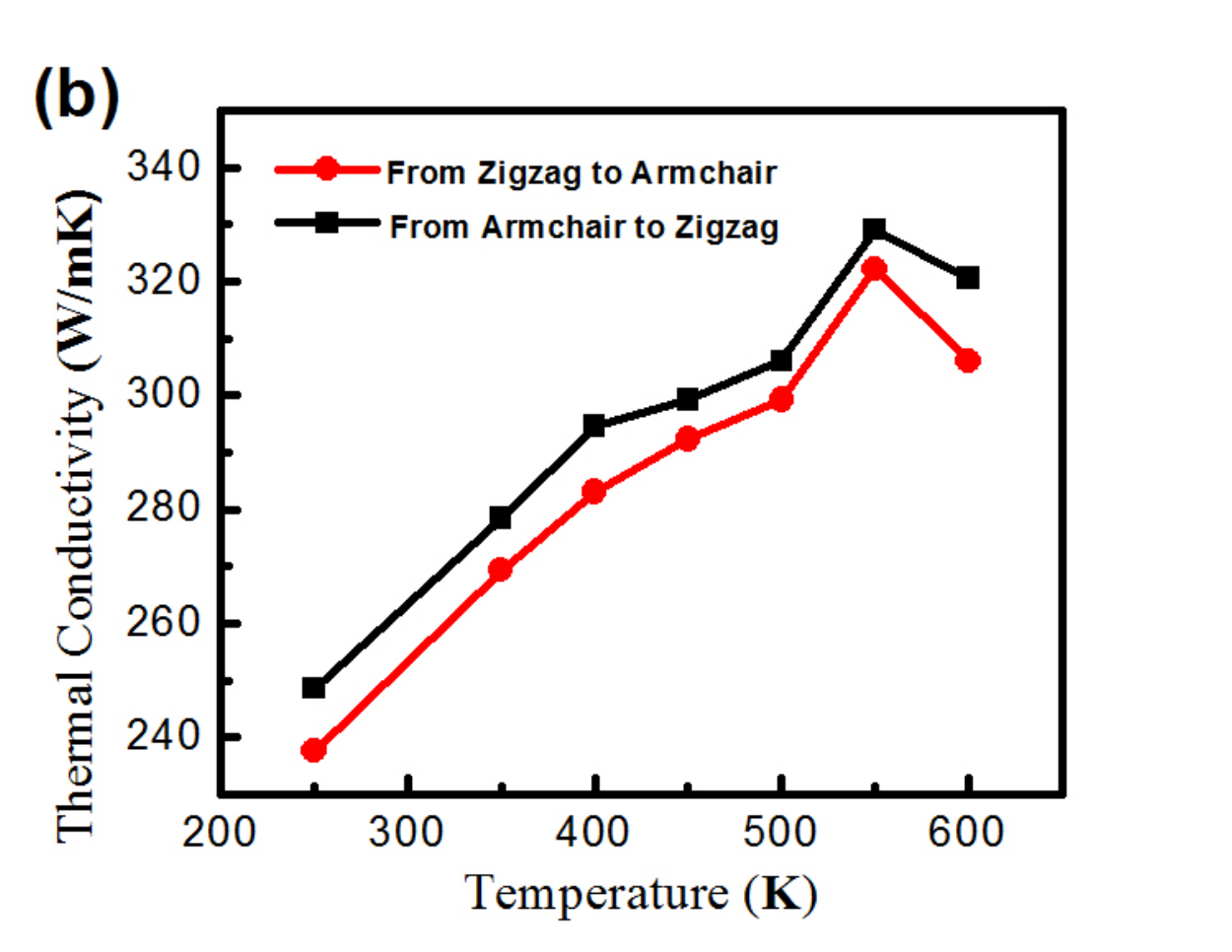}

\caption{The temperature dependence of (a) the Kapitza conductance and (b)
the thermal conductivity of armchair-zigzag asymmetric grain boundary.
Both the Kapitza conductance and the thermal conductivity increase
with the temperature increasing. The sudden drop of the Kapitza conductance
and the thermal conductivity is attributed to the changing of the
boundary structure at high temperature. In all the simulations, the
size of the sample is 10 nm.}
\end{figure}

\clearpage{}

\begin{figure}
\includegraphics[bb=30bp 0bp 360bp 278bp,scale=0.7]{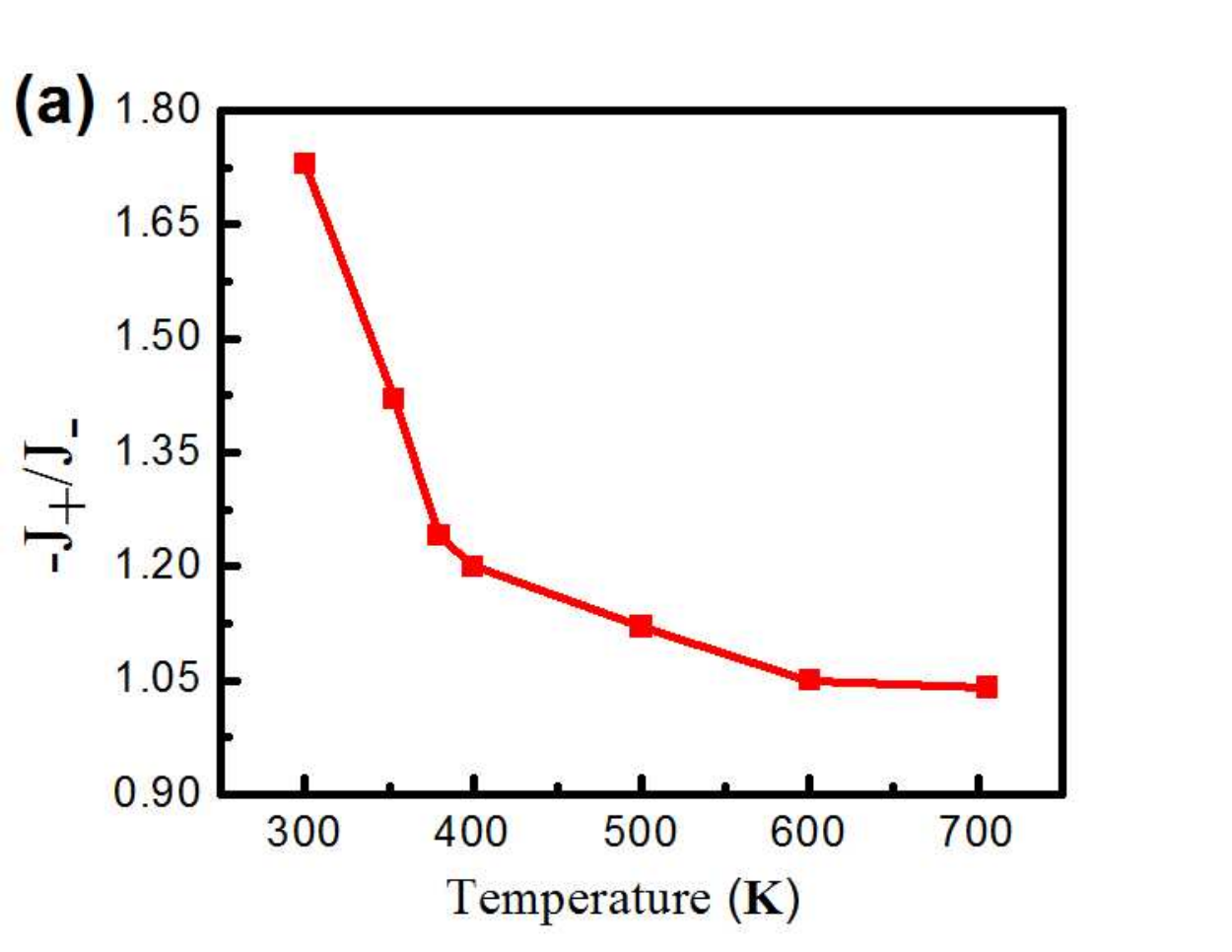}\includegraphics[scale=0.7]{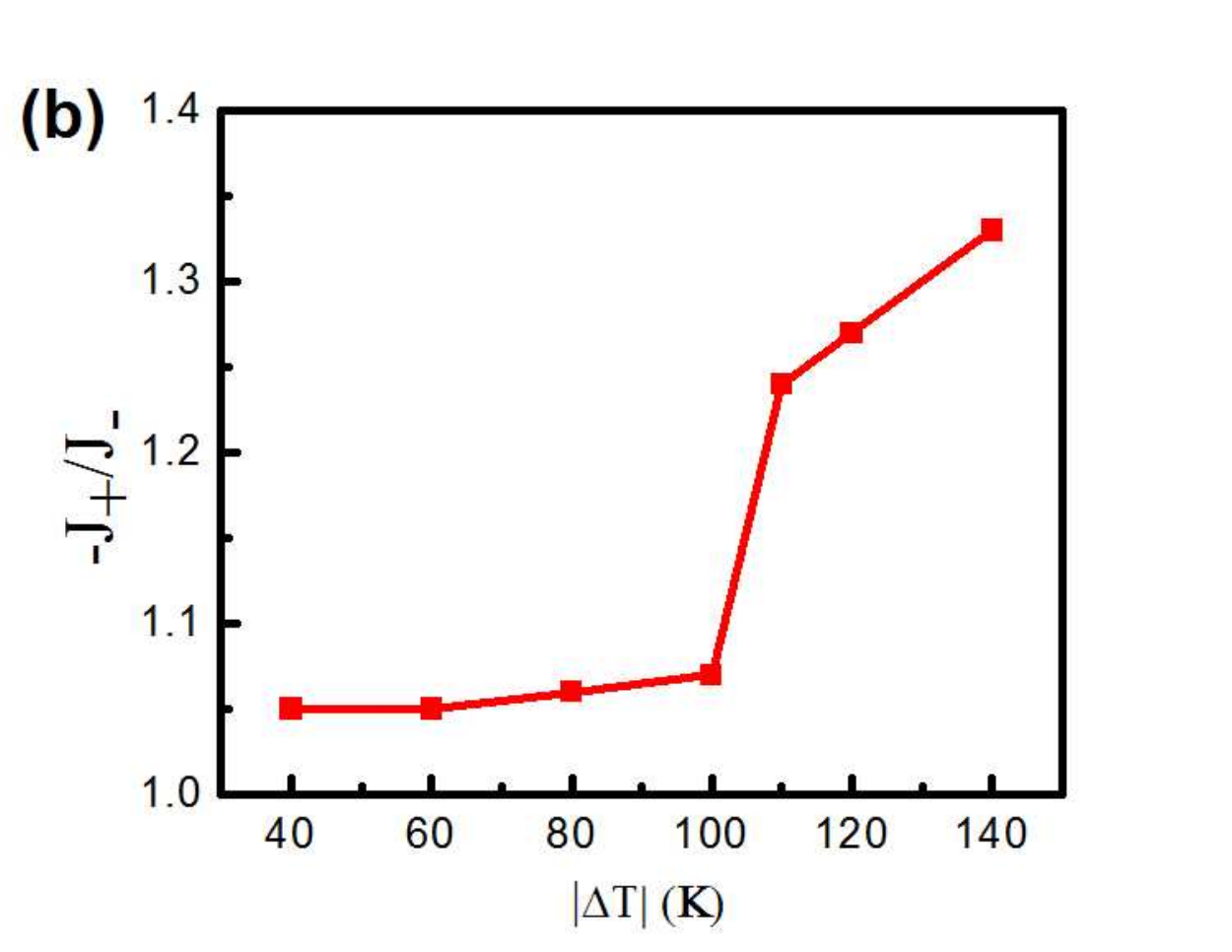}

\caption{The thermal rectification versus (a) different temperature and (b)
different temperature difference. In Figure 3(a) the temperature difference
is set to be 120 K. In Figure 3(b) the simulation temperature is set
to be 400 K. The effect of thermal rectification has been evaluated
by the ratio of the asymmetric heat flux along two reverse directions.
$J_{+}$ means the heat flux from the armchair part to the zigzag
part while $J_{-}$ means that the heat flux from the zigzag part
to the armchair part. In all the simulations, the size of the sample
is 10 nm.}
\end{figure}


\begin{thebibliography}{References}
\bibitem{1}A. K. Geim and K. S. Novoselov, Nat. Mater. \textbf{6},
183 (2007).

\bibitem{2}A. K. Geim, Science\textbf{ 324}, 1530 (2009).

\bibitem{3}T. J. Booth, P. Blake, R. R. Nair, D. Jiang, E. W. Hill,
U. Bangert, A. Bleloch, M. Gass, K. S. Novoselov, M. I. Katsnelson,
and A. K. Geim, Nano Lett. \textbf{8}, 2442 (2008).

\bibitem{4}C. Gomez-Navarro, R. T. Weitz, A. M. Bittner, M. Scolari,
A. Mews, M. Burghard, and K. Kern, Nano Letters \textbf{7}, 3499 (2007).

\bibitem{5}A. A. Balandin, S. Ghosh, W. Z. Bao, I. Calizo, D. Teweldebrhan,
F. Miao, and C. N. Lau, Nano Letters \textbf{8}, 902 (2008).

\bibitem{6}F. Wang, Y. B. Zhang, C. S. Tian, C. Girit, A. Zettl,
M. Crommie, and Y. R. Shen, Science \textbf{320}, 206 (2008).

\bibitem{7}K. S. Kim, Y. Zhao, H. Jang, S. Y. Lee, J. M. Kim, J.
H. Ahn, P. Kim, J. Y. Choi, and B. H. Hong, Nature \textbf{457}, 706
(2009).

\bibitem{8}A. Reina, X. T. Jia, J. Ho, D. Nezich, H. B. Son, V. Bulovic,
M. S. Dresselhaus, and J. Kong, Nano Lett. \textbf{9}, 30 (2009).

\bibitem{9}X. S. Li, W. W. Cai, J. H. An, S. Kim, J. Nah, D. X. Yang,
R. Piner, A. Velamakanni, I. Jung, E. Tutuc, S. K. Banerjee, L. Colombo,
and R. S. Ruoff, Science \textbf{324}, 1312 (2009).

\bibitem{10}W. W. Cai, A. L. Moore, Y. W. Zhu, X. S. Li, S. S. Chen,
L. Shi, and R. S. Ruoff, Nano Lett. \textbf{10}, 1645 (2010).

\bibitem{11}X. S. Li, C. W. Magnuson, A. Venugopal, J. H. An, J.
W. Suk, B. Y. Han, M. Borysiak, W. W. Cai, A. Velamakanni, Y. W. Zhu,
L. F. Fu, E. M. Vogel, E. Voelkl, L. Colombo, and R. S. Ruoff, Nano
Lett. \textbf{10}, 4328 (2010).

\bibitem{12}R. Grantab, V. B. Shenoy, and R. S. Ruoff, Science \textbf{330},
946 (2010).

\bibitem{13}O. V. Yazyev and S. G. Louie, Nat. Mater. \textbf{9},
806 (2010).

\bibitem{14}D. Gunlycke and C. T. White, Phys. Rev. Lett. \textbf{106},
136806 (2011).

\bibitem{15}J. Lahiri, Y. Lin, P. Bozkurt, Oleynik, II, and M. Batzill,
Nat. Nanotechnol. \textbf{5}, 326 (2010).

\bibitem{16}L. Z. Kou, C. Tang, W. L. Guo, and C. F. Chen, ACS Nano
\textbf{5}, 1012 (2011).

\bibitem{17}J. N. Hu, X. L. Ruan, and Y. P. Chen, Nano Lett. \textbf{9},
2730 (2009).

\bibitem{18}J. Haskins, A. Kinaci, C. Sevik, H. Sevincli, G. Cuniberti,
and T. Cagin, ACS Nano \textbf{5}, 3779 (2011).

\bibitem{19}J. N. Hu, S. Schiffli, A. Vallabhaneni, X. L. Ruan, and
Y. P. Chen, Appl. Phys. Lett. \textbf{97}, 133107(2010).

\bibitem{20}Akbar Bagri, Sang-Pil Kim, Rodney S. Ruoff, and Vivek
B. Shenoy, Nano Lett. \textbf{11}, 3917 (2011).

\bibitem{21}N. Yang, G. Zhang, and B. Li, Nano Lett. \textbf{65},
144306 (2002)

\bibitem{22}Z. X. Guo, D. Zhang, and X. G. Gong, Appl. Phys. Lett.
\textbf{95}, 163103 (2009).

\bibitem{23}Z. X. Guo, D. Zhang, Y. T. Zhai, and X. G. Gong, Nanotechnology
\textbf{21}, 285706 (2010).

\bibitem{24}Yuanyue Liu and Boris I. Yakobson, Nano Lett. \textbf{10},
2178 (2010).

\bibitem{25}Xiao-Fei Li, Ling-Ling Wang, Ke-Qiu Chen, and Yi Luo,
J. Chem. Phys. C \textbf{115}, 12616 (2011).

\bibitem{26}P. Koskinen, S. Malola; H. Hakkinen, Phys. Rev. Lett.
\textbf{101}, 115502 (2008).

\bibitem{27}Z. X. Guo and X. G. Gong, Front. Phys. China \textbf{4},
389 (2009).

\bibitem{28}S. Nos$\acute{e}$, J. Chem. Phys. \textbf{81}, 511 (1984)
W. G. Hoover, Phys. Rev. A \textbf{31}, 1695 (1985).

\bibitem{29}L. Lindsay and D. A. Broido, Phys. Rev. B \textbf{81},
205441 (2010).

\bibitem{30}L. Lindsay, D. A. Broido, and N. Mingo, Phys. Rev. B
\textbf{82}, 161402 (2010).

\bibitem{31}G. Wu and B. W. Li, Phys. Rev. B \textbf{76}, 085424
(2007).

\bibitem{32}Z. Y. Ong and E. Pop, Phys. Rev. B \textbf{84}, 075471
(2011).

\bibitem{33}Qing-Xiang Pei, Zhen-Dong Sha, and Yong-Wei Zhang, Carbon
\textbf{49}, 4752 (2011).

\bibitem{34}S. K. Chien, Y. T. Yang, and C. K. Chen, Appl. Phys.
Lett. \textbf{98}, 033107 (2011).

\bibitem{35}M. A. Angadi, T. Watanabe, A. Bodapati, X. C. Xiao, O.
Auciello, J. A. Carlisle, J. A. Eastman, P. Keblinski, P. K. Schelling,
and S. R. Phillpot, J. Appl. Phys. \textbf{99}, 114301 (2006).

\bibitem{36}P. K. Schelling, S. R. Phillpot, and P. Keblinski, J.
Appl. Phys. \textbf{95}, 6082 (2004).

\bibitem{37}A. Maiti, G. D. Mahan, and S. T. Pantelides, Solid State
Commun. \textbf{102}, 517 (1997).

\bibitem{38}V. Samvedi and V. Tomar, Nanotechnology \textbf{20},
365701 (2009).

\bibitem{39}Y. F. Chen, D. Y. Li, J. K. Yang, Y. H. Wu, J. R. Lukes,
and A. Majumdar, Physica B \textbf{349}, 270 (2004).

\bibitem{40}W. R. Zhong, M. P. Zhang, B. Q. Ai, and D. Q. Zheng,
Appl. Phys. Lett.\textbf{ 98}, 113107 (2011).

\bibitem{43}N. Yang, G. Zhang, and B. W. Li, Appl. Phys. Lett. \textbf{95},
033107 (2009).

\bibitem{44}N. Yang, G. Zhang, and B. W. Li, Appl. Phys. Lett. \textbf{93},
243111 (2008).

\bibitem{45}A. Rajabpour, S. M. V. Allaei, and F. Kowsary, Appl.
Phys. Lett. \textbf{99}, 051917 (2011).

\bibitem{46}B. W. Li, L. Wang, and G. Casati, Phys. Rev. Lett. \textbf{93},
184301 (2004).

\bibitem{47}T. Yamamoto, K. Watanabe, and K. Mii, Phys. Rev. B \textbf{70},
245402 (2004).

\bibitem{48}M. Yamada, Y. Yamakita, and K. Ohno, Phys. Rev. B\textbf{
77}, 054302 (2008).

\bibitem{49}N. A. Roberts and D. G. Walker, Int. J. Therm. Sci. \textbf{50},
648 (2011).\clearpage{} 
\begin{figure}
\includegraphics{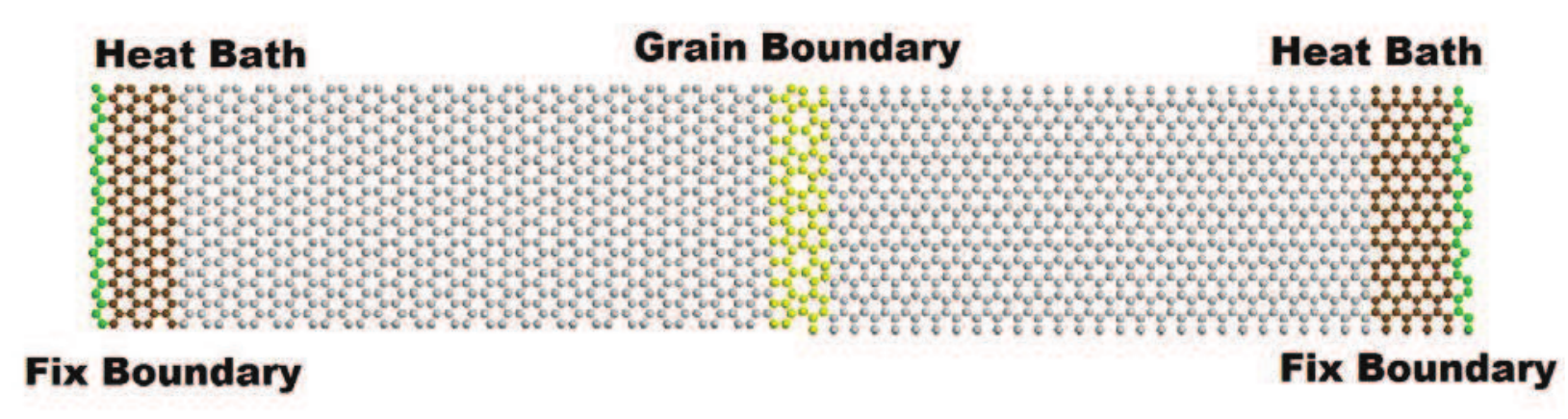} \caption{Structure of the asymmetric tilt grain boundary used in simulation.
The green atoms are set to be applied fix boundaries. The brown atoms
are set to be coupled to heat bath. The yellow atoms are the asymmetric
boundary structure composed of 5-pentagon and 7-heptagon topological
defects. The size of the sample is defined by the length between two
heat bath.}
\end{figure}


\end{thebibliography}
\end{document}